
\magnification=1200
\baselineskip=12pt

\rightline{UR-1347\ \ \ \ \ \ \ }
\rightline{ER-40685-796}

\baselineskip=18pt

\medskip

\centerline{\bf GELFAND-DIKII BRACKETS FOR NONSTANDARD LAX EQUATIONS}

\bigskip

\centerline{J.C. Brunelli}
\centerline{Ashok Das}

\medskip

\centerline{Department of Physics and Astronomy}
\centerline{University of Rochester}
\centerline{Rochester, NY 14627, USA}
\smallskip
\centerline{and}
\smallskip
\centerline{Wen-Jui Huang}
\centerline{Department of Physics}
\centerline{National Tsing Hua University}
\centerline{Hsinchu, Taiwan, Republic of China}

\bigskip
\bigskip

\centerline{\bf \underbar{Abstract}}

\medskip

We generalize the construction of Gelfand-Dikii brackets to the case of
nonstandard Lax equations.  We also discuss the possible origin of Kac-Moody
 algebras present in such systems.

\vfill\eject

\noindent {\bf I. \underbar{Introduction}:}

\medskip

Most integrable systems
 [1,2] can be understood in terms of pseudo-differential
 operators as follows [3,4].  Consider, for simplicity, the formal pseudo-
differential operator of the form
$$L_n = \partial^n + u_{-1} \partial^{n-1} + u_0 \partial^{n-2} +
\dots + u_{n-2} \eqno(1)$$
where `$\partial$' stands for the derivative operator ${\partial \over
\partial x}$ and the coefficient functions, $u_i (x,t)$, represent
dynamical variables.  Then,
$${\partial L_n \over \partial t_k} = \left[ \left( L_n^{k/n} \right)_+ , L_
n \right] = - \left[
\left( L_n^{k/n} \right)_- , L_n \right] \eqno(2)$$
with $k$ any positive integer which is not a multiple of $n$ defines a set
of evolution equations (nonlinear) for the dynamical variables,
$u_i (x,t)$, which are integrable.  Here $\left( L_n^{k/n} \right)_-$
$\left( \left( L_n^{k/n} \right)_+ \right)$ denotes the part of the
pseudo-differential operator $\left( L_n^{k/n} \right)$ containing only
negative (non negative) powers of $\partial$ and the system of equations is
conventionally known as the generalized KdV hierarchy.  The conserved
quantities (Hamiltonians) for the system can be seen to correspond to
(for more details see ref. 4)
$$\eqalign{H_k &= {n \over k}\ {\rm Tr}\ (L_n)^{k/n}\cr
\noalign{\vskip 4pt}%
&\equiv {n \over k}\ \int dx \ {\rm Res}\ \big( L_n^{k/n}\big)\cr}
 \eqno(3)$$
where Res (Residue) of a pseudo-differential operator is defined to be the
coefficient of $\partial^{-1}$.  These conserved charges can be easily
seen from Eqs. (2) and (3) to be in involution which is necessary for the
integrability of the system.

Given a pseudo-differential operator as in Eq. (1), one can define a dual
operator
$$Q_n = \partial^{-n} q_{-1} + \partial^{-n+1} q_0 + \dots +
\partial^{-1} q_{n-2} \eqno(4)$$
which allows us to define linear functionals of the dynamical variables,
 $u_i (x,t)$, as
$$F_{Q_n} (L_n) = \ {\rm Tr}\ L_n Q_n = \sum^{n-2}_{i=-1}
\int dx\ u_i (x) q_i (x) \eqno(5)$$
Given this, it can be shown that the Lax equation in Eq. (2) is Hamiltonian
with respect to the following Gelfand-Dikii brackets [3] (Hamiltonian
structures) for appropriate Hamiltonians in Eq. (3).
$$\eqalign{\big\{ F_{Q_n} (L_n) , F_{V_n} (L_n) \big\}_1 &= \ {\rm
 Tr}\ L_n [Q_n , V_n ]\cr
\big\{ F_{Q_n} (L_n) , F_{V_n} (L_n) \big\}_2 &= \ {\rm
 Tr}\big( \ Q_n L_n ( V_n L_n)_+ - L_n Q_n (L_n V_n )_+ )\cr
&= \ {\rm Tr}\ ((L_n (V_n L_n )_+ - ( L_n V_n )_+ L_n )Q_n )\cr}
\eqno(6)$$

If we set $u_{-1} = 0$ in Eq. (1) (which is consistent with the dynamical
equations in Eq. (2)), it can be easily checked that the presence or
absence of the term $\partial^{-n} q_{-1}$ in the dual makes no difference
in the definitions of $F_{Q_n} (L_n)$ and the first Hamiltonian structure.
However, the structure of the second Hamiltonian structure requires that $q
_{-1}$ be constrained for the Lax equation to be Hamiltonian with respect
to the second Hamiltonian structure [4].  Alternately, one can modify the
second Hamiltonian structure in such a case to be
$$\eqalign{\big\{ F_{Q_n} (L_n) , F_{V_n} (L_n) \big\}_2^\prime
 &= \ {\rm
 Tr}\ (Q_n L_n (V_n  L_n )_+ - L_n Q_n (L_n V_n )_+ )\cr
&\qquad + {1 \over n} \int dx\ \big( \int^x {\rm Res}\ [Q_n , L_n ] \big)
\ {\rm Res}\ [V_n , L_n ]\cr}\eqno(7)$$

There is a second class of integrable systems which have a nonstandard Lax
representation [5].  For simplicity, let us consider the equations describing
 dispersive water waves (conventionally, $J_0 = u,\ J_1 = h$
 and prime denotes derivative with respect to $x$)
$$\eqalign{{\partial J_0 \over \partial t} &= \big( 2 J_1 + J^2_0
- J^\prime_0 \big)^\prime\cr
\noalign{\vskip 4pt}%
{\partial J_1 \over \partial t} &= \big( 2 J_0 J_1 + J^\prime_1
\big)^\prime\cr}\eqno(8)$$
It is easy to check [5] that if we choose the pseudo-differential operator
$$L = \partial - J_0 + \partial^{-1} J_1$$
then Eq. (8) can be written in the Lax form
$${\partial L \over \partial t} = \left[ L , \left( L^2\right)_{\geq 1}
\right] \eqno(9)$$
where $\left( L^2 \right)_{\geq 1}$ stands for the purely differential part
of the pseudo-differential operator $L^2$ (with terms $a_n \partial^n ,\  n
\geq 1$). It is easy to check again that the conserved quantities of
the system
 (which are in involution) are given by
$$H_n = \ {\rm Tr}\ L^n = \int dx\ {\rm Res}\ L^n \qquad \qquad
n = 1,2,3,\dots \eqno(10)$$
The first few Hamiltonians can be directly read out from Eq. (10) to be
$$H_1 = \int dx\ J_1 \qquad H_2 = \int dx\ J_0 J_1 \qquad
H_3= \int dx\ \left( J^2_1 - J_0^\prime J_1 + J_1 J^2_0 \right) \eqno(11)$$
The set of equations (8) can be written in the Hamiltonian form with
the Poisson brackets
$$\eqalign{\big\{ J_0 (x) , J_0 (y) \big\}_1 &= 0 = \big\{ J_1 (x) , J_1
(y) \big\}_1\cr
\noalign{\vskip 4pt}%
\big\{ J_0 (x) , J_1 (y) \big\}_1 &= {\partial \over \partial x}\ \delta (x
-y) = \partial_x \delta (x-y)\cr}\eqno(12)$$
as
$$\eqalign{{\partial J_0 \over \partial t} &= \big\{ J_0 (x), H_3 \big\}_1
\cr
\noalign{\vskip 4pt}%
{\partial J_1 \over \partial t} &= \big\{ J_1 (x), H_3 \big\}_1\cr}
\eqno(13)$$
This defines the first Hamiltonian structure of the system.  However, we
note that with
$$\eqalign{\big\{ J_0 (x) , J_0 (y) \big\}_2 &=
 2 \partial_x \delta (x-y)\cr
\noalign{\vskip 4pt}%
\big\{ J_0 (x) , J_1 (y) \big\}_2 &= J_0 (y)
 \partial_x \delta (x-y) - \partial^2_x \delta (x-y)\cr
\noalign{\vskip 4pt}%
\big\{ J_1 (x) , J_0 (y) \big\}_2 &= J_0 (x)
 \partial_x \delta (x-y) + \partial^2_x \delta (x-y)\cr
\noalign{\vskip 4pt}%
\big\{ J_1 (x) , J_1 (y) \big\}_2 &= \big( J_1 (x) + J_1 (y)\big)
 \partial_x \delta (x-y)\cr}\eqno(14)$$
we can also write the equations in (8) as
$$\eqalign{{\partial J_0 \over \partial t} &= \big\{ J_0
(x) , H_2 \big\}_2\cr
\noalign{\vskip 4pt}%
{\partial J_1 \over \partial t} &= \big\{ J_1 (x) , H_2 \big\}_2\cr}
\eqno(15)$$
The Hamiltonian structure in Eq. (14) defines the second structure of the
theory.  While these Hamiltonian structures have been obtained
directly from the structure of the equations of motion as well as from
 other methods [6-9], a Gelfand-Dikii construction for them is so far lacking.
In this letter, we would generalize the Gelfand-Dikii brackets to
nonstandard Lax equations and bring out some connections with the
Kac-Moody algebras and covariant Lax operators.

In closing this section, we note that the set of equations (8)
 can be mapped on to the nonlinear Schr\"odinger equation with
 the identifications
$$\eqalign{J_0 &= - {\psi^\prime \over \psi}\cr
J_1 &= \overline \psi \psi\cr}\eqno(15^\prime)$$
and appropriate scaling of the variables $x$ and $t$.  Recently,
there has been a lot of interest in this system [6-10] from
various points of view.  The construction of the Gelfand-Dikii
brackets for such systems is, therefore, of significance.

\medskip

\noindent {\bf II. \underbar{Gelfand-Dikii Brackets}:}

\medskip

Given the pseudo-differential operator
$$\eqalign{L &= \partial - J_0 + \partial^{-1} J_1\cr
&= \partial + \overline J_0 + \partial^{-1} \overline J_1
\qquad\qquad \big( \overline J_0 = - J_0 \quad \overline J_1 = J_1 \big)
\cr}\eqno(16)$$
let us define the dual as
$$Q = q_0 + q_1 \partial^{-1} \eqno(17)$$
We can now define a linear functional of $\left( \overline J_0 ,
\overline J_1 \right)$ as
$$F_Q (L) = \ {\rm Tr}\ LQ = \sum^1_{i=0}
\int dx\ q_{1-i} (x)
\overline J_i (x) \eqno(18)$$
It is now straightforward to check that
$$\left\{ F_Q (L), F_V (L) \right\}_1 =\
{\rm Tr}\ L [Q,V]$$
does, indeed, reproduce the first Hamiltonian structure in Eq. (12).
However, the definitions of the brackets for the second structure in Eq.
(6) or (7) fail to give the Hamiltonian structure in Eq. (14).  On the
other hand, it is clear that we are dealing here with a nonstandard Lax
representation and consequently, the definitions of the Gelfand-Dikii
brackets in Eqs. (6) and (7) may need to be modified.

To understand the structure of the Gelfand-Dikii brackets in the present
case, let us note that with
$$\partial \omega = J_0 = - \overline J_0 \eqno(19)$$
we have
$$\eqalign{e^{- \omega} \partial e^\omega &= \partial + J_0 =
\partial - \overline J_0 \equiv D\cr
 e^{-\omega} \partial^n e^\omega &= D^n = \big( \partial + J_0 \big)^n
= \big( \partial - \overline J_0 \big)^n\cr}\eqno(20)$$
where $n$ is any integer, positive or negative.  Given this, we note that
$$\eqalign{L &= \partial - J_0 + \partial^{-1} J_1 \cr
&= e^\omega \hat L e^{- \omega}\cr}\eqno(21)$$
where
$$\hat L = D + \overline J_0 + D^{-1} \overline J_1
= D-J_0 + D^{-1} J_1 \eqno(22)$$
It is now straightforward to check that
$${\partial \hat L \over \partial t} = \left[ \hat L,
 \left( \left( \hat L\right)^2
\right)_+  \right] \eqno(23)$$
gives
$$D^{-1} {\partial J_1 \over \partial t} - D^{-1} {\partial J_0 \over
\partial t} \ D^{-1} J_1 = D^{-1} \left( 2 J_0 J_1 + J^\prime_1
\right)^\prime - D^{-1} \left( 2 J_1 + J^2_0 - J^\prime_0 \right)^\prime
D^{-1} J_1 \eqno(24)$$
Comparing, we see that these give the same equations as in Eq. (8).  Thus,
we note that the equations (8) can be written in a
\underbar{standard Lax form} with
the pseudo-differential operator $\hat L$ which is expanded in the basis $D$.
The Gelfand-Dikii brackets would now follow.  Let us define the dual as
$$\hat Q = q_0 + q_1 D^{-1} = e^{- \omega} Q e^\omega \eqno(25)$$
so that
$$F_{\hat Q} (\hat L) =\ {\rm Tr}\ \hat L \hat Q =\
{\rm Tr}\ LQ = F_Q (L) = \sum^1_{i=0} \int dx\
q_{1-i} (x) \overline J_i (x) \eqno(26)$$
It follows now that
$$\left\{ F_{\hat Q} (\hat L) , F_{\hat V} (\hat L )\right\}_1 =\
{\rm Tr}\ \hat L [ \hat Q , \hat V]$$
would lead to the first Hamiltonian structure of the system.  Using
Eqs. (21), (25) and (26) it follows then that
$$\left\{ F_Q (L) , F_V (L) \right\}_1 =\
{\rm Tr}\ L[Q,V] \eqno(27)$$
would lead to the first Hamiltonian structure for the system as we have
noted earlier.

The second Hamiltonian structure for a standard Lax representation is
defined to be (see Eq. (6))
$$\left\{ F_{\hat Q} (\hat L), F_{\hat V} (\hat L)\right\}_2 =
 \ {\rm Tr}\ \left( \left( \hat L \left( \hat V \hat L \right)_+
- \left( \hat L \hat V \right)_+ \hat L \right) \hat Q \right) \eqno(28)$$
However, a little analysis reveals that $\hat L (\hat V \hat L)_+ - (\hat L
\hat V)_+ \hat L$ is a pseudo-differential operator of leading order
zero. From Eqs. (23) and (24), it then follows that this definition cannot
describe the Lax equation as a Hamiltonian equation.  (This is similar to
the constrained case $u_{-1} = 0$ we discussed earlier.
 See ref. 4 for details.) To remedy this, we
define
$$\eqalign{\overline{\hat Q} &= q_{-1} D + q_0 + q_1 D^{-1} = q_{-1} D +
\hat Q\cr
\overline{\hat V} &= v_{-1} D + v_0 + v_1 D^{-1} = v_{-1} D + \hat V\cr}
\eqno(29)$$
and note that
$$F_{\overline{\hat Q}} (\hat L) = F_{\hat Q} (\hat L) =
F_Q (L) \eqno(30)$$
However, we observe that $( \hat L (\overline{\hat V} \hat L)_+ -
(\hat L \overline{\hat V})_+ \hat L)$ will now be an operator of leading
order $-1$ provided
$$v_{-1} J_1 = v_1 \eqno(31)$$
In other words, the system becomes a constrained one similar to the case of
generalized KdV hierarchy which we have discussed earlier.  The appropriate
definition of the second Hamiltonian structure is, then,
$$\left\{ F_{\hat Q} (\hat L) , F_{\hat V} (\hat L) \right\}_2 =\
{\rm Tr}\ \left( \left( \hat L \left( \overline{\hat V}
\hat L \right)_+ -
\left( \hat L \overline{\hat V}\right)_+
\hat L \right) \overline{\hat Q} \right) \eqno(32)$$
with constraints
 on $q_{-1}$ and $v_{-1}$ of the type in
 Eq. (31) and it is straightforward to check that this
indeed reproduces the second Hamiltonian structure of Eq. (14).
Transforming this back to the original variables we see that with
$$\eqalign{
\overline Q &= q_{-1} \partial + Q\cr
\overline V &= v_{-1} \partial + V\cr} \eqno(33)$$
where
$$\eqalign{q_{-1} J_1 &= q_1\cr
v_{-1} J_1 &= v_1\cr} \eqno(34)$$
the second Hamiltonian structure can be obtained from
$$\left\{ F_Q (L) , F_V (L) \right\}_2 =\
{\rm Tr}\ \left( \left( L \left( \overline V L \right)_+ -
\left( L \overline V \right)_+ L \right) \overline Q \right) \eqno(35)$$

It is interesting to note that while the definition of the dual as in Eq.
(29) or (33) does not change the definition of the linear functional (see
Eq. (30)), it affects the structure of the first Hamiltonian structure.
(This is different from the case of the generalized KdV hierarchy.)
  It
is, therefore, more appropriate to leave the definition of the dual
 unchanged in the
present case and modify, instead, the definition of the second bracket to be
$$\eqalign{\big\{ F_{\hat Q} (\hat L), F_{\hat V} (\hat L)\big\}_2
=\ &{\rm Tr}\ \big( \big( \hat L ( \hat V \hat L)_+ - (\hat L \hat V)_+
\hat L) \hat Q \big)\cr
&+ \int dx\ \bigg[ \big( \int^x {\rm Res}\ [\hat Q , \hat L ]\big)\
{\rm Res}\ [\hat V , \hat L]\cr
&+ \ {\rm Res}\ [\hat Q , \hat L] \ {\rm Res}\ \big( D^{-1} \hat L \hat V
\big) -\ {\rm Res}\ [\hat V , \hat L] \ {\rm Res}\
\big( D^{-1} \hat L \hat Q \big) \bigg]\cr}\eqno(36)$$
In terms of the original variables, this would then give
$$\eqalign{\big\{ F_Q (L), F_V (L)\big\}_2 = \ &{\rm Tr}\ ( (
L(VL)_+ - (LV)_+ L)Q )\cr
&+ \int dx\ \bigg[ \bigg( \int^x \ {\rm Res}\ [Q,L] \bigg) \ {\rm Res}\ [V,L]
\cr
&+\ {\rm Res}\ [Q,L] \ {\rm Res}\ \big( \partial^{-1} LV \big) -\ {\rm
 Res}\ [V,L] \ {\rm Res}\
\big( \partial^{-1} LQ \big) \bigg]\cr}\eqno(37)$$
These brackets are by definition antisymmetric and give the Hamiltonian
structure in Eq. (14) which we know to satisfy the Jacobi identity.

\medskip

\noindent {\bf III. \underbar{Kac-Moody Algebras}:}

\medskip

The second Hamiltonian structure of integrable systems, in general,
corresponds to interesting, nontrivial symmetry algebras.  Thus, for
example, the Virasoro algebra arises as the second Hamiltonian structure of
the KdV equation [11] whereas the Boussinesq equation has the $W_3$-algebra
as its second Hamiltonian structure [12] and so on.  In the
present case, we note that while
$$\{ J_0 (x) , J_0 (y) \}_2 = 2 \partial_x \delta (x-y)$$
represents the $U(1)$ current algebra, the entire second Hamiltonian
structure in Eq. (14) does not appear familiar.  We note, however, that if
we define
$$\eqalign{J(x) &= J_0 (x)\cr
T(x) &= J_1 (x) - {1 \over 2}\ J^\prime_0 (x)\cr}\eqno(38)$$
then, the algebra in terms of these variables becomes
$$\eqalign{\{ J(x) , J(y) \}_2 &= 2 \partial_x \delta (x-y)\cr
\{ T(x) , J(y) \}_2 &= J(x) \partial_x \delta (x-y)\cr
\{ T(x) , T(y)\}_2 &= (T (x) + T(y)) \partial_x \delta (x-y) +
{1 \over 2}\ \partial^3_x \delta (x-y)\cr}\eqno(39)$$
We recognize this as the Virasoro-Kac-Moody algebra for a $U(1)$ current
 [13,14].
Since both the KdV and the nonlinear Schr\"odinger equation can be
 obtained from a zero curvature condition associated with SL(2,{\bf R}), a
hidden Virasoro algebra in the nonlinear Schr\"odinger equation may have been
expected.  It is, however, interesting to find a Virasoro-Kac-Moody
 algebra hidden in the nonlinear Schr\"odinger equation (see Eq.
(15$^\prime$)).

We also note that we can define a new spin-2 generator as
$$W (x) = T(x) - {1 \over 4}\  J^2 (x)
= J_1 (x) - {1 \over 2}\ J^\prime (x) -
{1 \over 4}\ J^2 (x) \eqno(40)$$
and it would satisfy the
Virasoro subalgebra.  However,  with this redefinition,
it is easy to see that the algebra (39) decouples to
$$\eqalign{\{ J(x), J(y)\}_2 &= 2 \partial_x \delta (x-y)\cr
\{ W(x), J(y) \}_2 &= 0\cr
\{ W(x) , W(y) \}_2 &=
\big( W(x) + W(y) \big) \partial_x \delta (x-y) +
{1 \over 2}\ \partial^3_x \delta (x-y)\cr}\eqno(41)$$
It is interesting to note that the definition in Eq. (40) is a Miura
transformation which maps the KdV equation to the mKdV equation [15].  The
Hamiltonian structure in Eq. (41) can again be thought of as a direct sum
of the Hamiltonian structures for the mKdV and the KdV systems.  However,
we note that while the integral of $T\ (=H_1)$ is a conserved quantity of
the system, integral of $W$ is not conserved.  Consequently, it is more
natural to discuss the algebra in the basis of $J$ and $T$ [14].

Kac-Moody algebras are known to arise within the context of
pseudo-differential operators when the basis of expansion is a gauge
covariant derivative [13,14].  We do not fully understand yet the origin of the
Kac-Moody algebras in the present system.  However, we offer the following
observations as providing a possible connection.  Note that we can write
$$\eqalign{L &= \partial - J_0 + \partial^{-1} J_1\cr
&= \partial^{-1} \big( \partial^2 - \partial J_0 + J_1 \big)\cr
&= \partial^{-1} \big( {\cal D}^2 + W \big) =
\partial^{-1} {\cal L}\cr}\eqno(42)$$
where we have defined
$$\eqalign{{\cal D} &= \partial - {1 \over 2}\ J_0 \cr
\noalign{\vskip 4pt}%
W &= J_1 - {1 \over 2}\ J^\prime_0 - {1 \over 4}\ J^2_0\cr
\noalign{\vskip 4pt}%
{\cal L} &= {\cal D}^2 + W\cr}\eqno(43)$$
We can think of ${\cal D}$ as a covariant derivative and consequently,
the pseudo-differential operator ${\cal L}$ can be thought of as a
covariant generalization of the Lax operator for the KdV system which is
known [13,14] to lead to a Virasoro-Kac-Moody algebra.  However, more work is
needed to understand the connection with the Kac-Moody algebras fully.

\medskip

\noindent {\bf IV. \underbar{Conclusion}:}

\medskip

We have proposed a generalization of the Gelfand-Dikii brackets to the case
of nonstandard Lax equations.  Although our discussion has been entirely
within the context of the dispersive water wave equations
(which can also be mapped to the
nonlinear Schr\"odinger equation), the method is
quite general and would extend readily to other nonstandard equations.  We
have tried to bring out the hidden Virasoro-Kac Moody algebra in the second
Hamiltonian structure of the system and have offered some partial
explanation for its origin.  However, more work is needed for a deeper
understanding of this question.

This work was supported in part by U.S. Department of Energy Grant No.
DE-FG-02-91ER40685. J.C.B. would like to thank CNPq, Brazil for
financial support.  W.-J.H. is supported by the National Science Council
of the Republic of China under Grant No. NSC-83-0208-M-007-008.

\vfill\eject

\noindent {\bf \underbar{References}}

\medskip

\item{1.} L.D. Faddeev and L.A. Takhtajan, ``Hamiltonian Methods in the
Theory of Solitons", Springer, Berlin (1987).

\item{2.} A. Das, ``Integrable Models", World Scientific (1989).

\item{3.} I.M. Gelfand and L.A. Dikii, Funct. Anal. Appl. {\bf 11},
93 (1977).

\item{  } M. Adler, Invent. Math. {\bf 50}, 219 (1979).

\item{  } L.A. Dickey, ``Soliton Equations and Hamiltonian Systems",
 World Scientific (1991).

\item{4.} A. Das and W.J. Huang, J. Math. Phys. {\bf 33}, 2487
 (1992) and references therein.

\item{5.} B.A. Kupershmidt, Comm. Math. Phys. {\bf 99}, 51 (1985).

\item{6.} H. Aratyn, L.A. Ferreira, J. F. Gomes and A. H. Zimerman, Nucl.
 Phys. {\bf B402}, 85 (1993); H. Aratyn, L.A. Ferreira,
J.F. Gomes and A.H. Zimerman, hep-th/9304152.

\item{7.} H. Aratyn, E. Nissimov and S. Pacheva, Phys. Lett.
{\bf B314}, 41 (1993).

\item{8.} L. Bonora and C. S. Xiong, Phys. Lett. {\bf B285}, 191 (1992).

\item{9.} L. Bonora and C.S. Xiong, Int. J. Mod. Phys. {\bf A8}, 2973
 (1993).

\item{10.} M.D. Freeman and P. West, Phys. Lett. {\bf B295}, 59 (1992).

\item{11.} F. Magri, J. Math. Phys. {\bf 19}, 1156 (1978).
\item{   } J.L. Gervais, Phys. Lett. {\bf B160}, 277 (1985).

\item{12.} H.P. McKean, Adv. Math. Supp. {\bf 3}, 217 (1978).

\item{  } P. Mathieu, Phys. Lett. {\bf 208B}, 101 (1988).

\item{  } Q. Wang, P.K. Panigrahi, U. Sukhatme and Y. Keung, Nucl. Phys.
 {\bf B344}, 196 (1990).

\item{  } A. Das and S. Roy, Int. J. Mod. Phys. {\bf A6}, 1429 (1991).

\item{13.} A. Das and S. Roy, J. Math. Phys. {\bf 32}, 869 (1991).

\item{14.} W.J. Huang, J. Math. Phys. {\bf 35}, 993 (1994).

\item{15.} R.M. Miura, J. Math. Phys. {\bf 9}, 1202 (1968).

\end